\documentclass[12pt]{article}
\usepackage{OSAmeet}
\usepackage{times}
\usepackage{epsfig}
\usepackage{setspace}
\usepackage{overcite}
\begin{document}                
\title{Optical interpretation of special relativity and quantum mechanics }
\author{Jos\'e B. Almeida }
\address{Universidade do Minho, Physics Department, 4710-057 Braga,
Portugal }
 \email{Tel: +351-253 604390, e-mail: bda@fisica.uminho.pt}
%


\HeaderAuthorTitleMtg{Almeida}{Optical
interpretation$\ldots$}{OSA Annual meeting}
%
\begin{abstract}                
The present work shows that through a suitable change of
variables relativistic dynamics can be mapped to light
propagation in a non-homogeneous medium. A particle's trajectory
through the modified space-time is thus formally equivalent to a
light ray and can be derived from a mechanical equivalent of
Fermat's principle. The similarities between light propagation
and mechanics are then extended to quantum mechanics, showing
that relativistic quantum mechanics can be derived from a wave
equation in modified space-time. Non-relativistic results, such
as de Broglie's wavelength, Schr\"odinger equation and
uncertainty principle are shown to be direct consequences of the
theory and it is argued that relativistic conclusions are also
possible.
\end{abstract}
\section{Introduction}
This paper is presented in a rather crude state; the text is
imperfect and some conclusions are deferred to ulterior
publications; nevertheless the author feels that this work must
be diffused even in a preliminary stage due to its significance.
The author's presence at the OSA's annual meeting provided an
opportunity for the presentation of his work that he could not
despise.

The similarities between light propagation and wave mechanics
have been pointed out by numerous authors, although a perfect
mapping from one system to the other has never been achieved.
Almeida et al. \cite{Almeida00:3} showed that near-field light
diffraction could be calculated using the Wigner Distribution
Function (WDF) and obtained results proving the existence of
super-resolution in certain circumstances.

The study of wide angle light propagation makes use of a
transformation which brings to mind the Lorentz transformation of
special relativity. It was then natural to try an association of
Newtonian mechanics to paraxial optics and special relativity to
wide angle propagation. This process promoted the definition of a
coordinate transformation to render the relativistic space
homologous to the optical space. The introduction of a
variational principle allowed the derivation of relativistic
dynamics in the modified space-time in a process similar to the
derivation of optical propagation from Fermat's principle. One
important consequence is that each particle travels through
modified space-time with the speed of light.

The similarity could be carried further to diffraction phenomena
and quantum mechanics. It was postulated that a particle has an
intrinsic frequency related to its mass and many important
results were derived directly from this statement. More general
results will probably be feasible in the future.
\section{Notes on Hamiltonian optics}
The propagation of an optical ray is governed by Fermat's
principle, which can be stated \cite{Born80}:
\begin{equation}
 \label{eq:Fermat}
 \delta \int_{P_1}^{P_2} n d s = 0~.
\end{equation}
The integral quantity is called \emph{point characteristic} and
measures the optical path length between points $P_1$ and $P_2$.
\begin{equation}
 \label{eq:pointchar}
 V(x_1,y_1,x_2,y_2) = \int_{P_1}^{P_2} n d s ~.
\end{equation}

The quantity $d s$ is the length measured along a ray path and
can be replaced by:
\begin{equation}
 \label{eq:ds}
 d s = \frac{dz}{(1 - u_x^2 - u_y^2)^{1/2}}= \frac {dz}{u_z}~,
\end{equation}
where $u_x$, $u_y$ and $u_z$ are the ray direction cosines with
respect to the $x$, $y$ and $z$ axes.

Inserting Eq.\ (\ref{eq:ds}) into Eq.\ (\ref{eq:Fermat}) we have:
\begin{eqnarray}
 && \delta \int_{z_1}^{z_2} \frac{n dz} {(1 - u_x^2 - u_y^2)^{1/2}} = 0 ~,\nonumber \\
 && \delta \int_{z_1}^{z_2} \frac{n dz} {u_z} = 0 ~.
\end{eqnarray}

Of course we can also write:
\begin{equation}
 \label{eq:dsslope}
 d s = (1 + \dot{x}^2 + \dot{y}^2)^{1/2} dz~,
\end{equation}
with
\begin{equation}
 \label{eq:slopes}
 \dot{x} = \frac{d x}{d z}~~\mathrm{and}~~
 \dot{y} = \frac{d y}{dz}~.
\end{equation}

It is easy to relate $\dot{x}$ and $\dot{y}$ to $u_x$ and
 $u_y$:
\begin{eqnarray}
 \label{eq:uv}
 u_x &=& \frac{\dot{x}}{(1 + \dot{x}^2+ \dot{y}^2)^{1/2}} \nonumber ~,\\
 u_y &=& \frac{\dot{y}}{(1 + \dot{x}^2+ \dot{y}^2)^{1/2}}~,
\end{eqnarray}
where only the positive root is considered.

Inserting Eq. (\ref{eq:dsslope}) into Eq.\ (\ref{eq:Fermat}) we
get:
\begin{equation}
 \delta \int_{z_1}^{z_2} n (1 + \dot{x}^2 + \dot{y}^2)^{1/2} dz = 0~.
\end{equation}

 We can use the position coordinates $x$ and $y$ as generalized
coordinates and $z$ for time, in order to define the Lagrangian.
We have \cite{Goldstein80, Jose98}:
\begin{equation}
 \label{eq:Lagrangian}
 L = n (1 + \dot{x}^2 + \dot{y}^2)^{1/2}
\end{equation}

Euler Lagrange's propagation equations are:
\begin{eqnarray}
 \label{eq:lagrange}
 \frac{d}{dz} \frac{\partial L}{\partial \dot{x}} - \frac{\partial L}{\partial x} &=& 0 ~, \\
 \frac{d}{dz} \frac{\partial L}{\partial \dot{y}} - \frac{\partial L}{\partial y} &=& 0 ~;
\end{eqnarray}

We can go a step further if we define a system Hamiltonian and
write the canonical equations; we start by finding the components
of the conjugate momentum ($\mathbf{p}$) from the Lagrangian.
Knowing that $n$ is a function of $x$, $y$ and $z$, the conjugate
momentum components can be written as:
\begin{eqnarray}
 \label{eq:momenta}
 p_x &=& \frac{\partial L}{\partial \dot{x}} = \frac{n \dot{x}}{(1 + \dot{x}^2 + \dot{y}^2)^{1/2}} ~,
 \nonumber \\
 p_y &=& \frac{\partial L}{\partial \dot{y}} = \frac{n \dot{y}}{(1 + \dot{x}^2 + \dot{y}^2)^{1/2}} ~.
\end{eqnarray}
If we consider Eq.\ (\ref{eq:uv}), the result is:
\begin{eqnarray}
 \label{eq:momenta2}
 p_x &=& n u_x ~, \nonumber \\
 p_y &=& n u_y ~.
\end{eqnarray}

The system Hamiltonian is:
\begin{eqnarray}
 \label{eq:Hamiltonian}
 H &=& p_x \dot{x} + p_y \dot{y} - L \nonumber \\
   &=& \frac{n (\dot{x}^2 + \dot{y}^2)}{(1 + \dot{x}^2 + \dot{y}^2)^{1/2}}
 - L \nonumber \\
   &=& \frac{-n }{(1 + \dot{x}^2 + \dot{y}^2)^{1/2}} \nonumber \\
   &=& -n (1 - u_x^2 -u_y^2)^{1/2}~, \nonumber \\
   &=& -n u_z ~,
\end{eqnarray}

 The Hamiltonian has the interesting property of having the
dependence on the generalized coordinates and time, separated
from the dependence on the conjugate momentum. The canonical
equations are:
\begin{eqnarray}
 \label{eq:canonical}
   \dot{x}&=& \frac{\partial H}{\partial p_x} = \frac{u_x}{(1-u_x^2-u_y^2)^{1/2}}~ =~ \frac{u_x}{u_z}~, \nonumber \\
   \dot{y}&=& \frac{\partial H}{\partial p_y} = \frac{u_y}{(1-u_x^2-u_y^2)^{1/2}}~ =~ \frac{u_y}{u_z}~, \nonumber \\
   n \dot{u}_x+ \dot{n} u_x &=& -\frac{\partial H}{\partial x} = (1-u_x^2-u_y^2)^{1/2} \frac{\partial n}{\partial x}
   ~=~ u_z \frac{\partial n}{\partial x}~, \nonumber \\
   n \dot{u}_y+ \dot{n} u_y &=& -\frac{\partial H}{\partial y} = (1-u_x^2-u_y^2)^{1/2} \frac{\partial n}{\partial y}
   ~=~ u_z \frac{\partial n}{\partial y}~.
\end{eqnarray}
Obviously the first two canonical equations represent just a
trigonometric relationship.

It is interesting to note that if the refractive index varies
only with $z$, then the conjugate momentum will stay unaltered;
the direction cosines will vary accordingly to keep constant the
products $n u_x$ and $n u_y$.

We will now consider an non-homogeneous medium with a direction
dependent refractive index and will add this dependence as a
correction to a nominal index.
\begin{equation}
    \label{eq:indexdiff}
    n = n_0 - \frac{n_c}{\left(1+ \dot{x}^2 + \dot{y}^2 \right)^{1/2}},
\end{equation}
where $n_0$ is the nominal index and $n_c$ is a correction
parameter. Eq.\ (\ref{eq:Lagrangian}) becomes
\begin{equation}
    \label{eq:Lagnoni}
    L = n_0 \left(1 + \dot{x}^2 + \dot{y}^2 \right) - n_c.
\end{equation}

We will follow the procedure for establishing the canonical
equations in this new situation. It is clear that the momentum is
still given by Eq.\ (\ref{eq:momenta2}) if $n$ is replaced by
$n_0$.

The new Hamiltonian is given by
\begin{equation}
    \label{eq:Hamiltoniandiff}
    H = -n_0 (1-u_x^2-u_y^2) + n_c,
\end{equation}
and the canonical equations become
\begin{eqnarray}
 \label{eq:canonicaldiff}
   \dot{x}&=& \frac{\partial H}{\partial p_x} = \frac{u_x}{(1-u_x^2-u_y^2)^{1/2}}~ =~ \frac{u_x}{u_z}~, \nonumber \\
   \dot{y}&=& \frac{\partial H}{\partial p_y} = \frac{u_y}{(1-u_x^2-u_y^2)^{1/2}}~ =~ \frac{u_y}{u_z}~, \nonumber \\
   n_0 \dot{u}_x+ \dot{n}_0 u_x &=& -\frac{\partial H}{\partial x} =
    u_z \frac{\partial n_0}{\partial x}~-\frac{\partial n_c}{\partial x}~, \nonumber \\
   n_0 \dot{u}_y+ \dot{n}_0 u_y &=& -\frac{\partial H}{\partial y} =
   u_z \frac{\partial n_0}{\partial y}~-\frac{\partial n_c}{\partial y}~.
\end{eqnarray}

The present discussion of non-homogeneous media is not completely
general but is adequate for highlighting similarities with
special relativity and quantum mechanics, as is the purpose of
this work.
\section{Diffraction and Wigner distribution function}
Almeida et al. \cite{Almeida00:3} have shown that the high
spatial frequencies in the diffracted spectrum cannot be
propagated and this can even, in some cases, lead to a
diffraction limit much lower than the wavelength; here we detail
those arguments.

The Wigner distribution function (WDF) of a scalar, time
harmonic, and coherent field distribution $\varphi(\mathbf{q},z)$
can be defined at a $z=\mathrm{const.}$ plane in terms of either
the field distribution or its Fourier transform
$\overline{\varphi}(\mathbf{p})=\int
\varphi(\mathbf{q})\exp(-ik\mathbf{q}^T \mathbf{p})d\mathbf{q}$
\cite{Bastiaans79, Dragoman97, Bastiaans97}:
\begin{eqnarray}
 \label{eq:Wigner}
 W(\mathbf{q},\mathbf{p})&=&\int
 \varphi\left(\mathbf{q}+\frac{\mathbf{q}'}{2}\right)
 \varphi^*\left(\mathbf{q}-\frac{\mathbf{q}'}{2}\right)
 \exp\left(-i k \mathbf{q}'^T \mathbf{p}\right)d \mathbf{q}' \\
 &=& \frac{k^2}{4\pi^2}\int
 \overline{\varphi}\left(\mathbf{p}+\frac{\mathbf{p}'}{2}\right)
 \overline{\varphi}^*\left(\mathbf{p}-\frac{\mathbf{p}'}{2}\right)
 \exp\left(i k \mathbf{q}^T \mathbf{p}'\right)d \mathbf{p}'~,
\end{eqnarray}
 where $k=2 \pi/
\lambda$, $^*$ indicates complex conjugate and
\begin{eqnarray}
    \label{eq:vectors}
    \mathbf{q} &=& \left(x,y \right), \\
    \label{eq:vectors1}
    \mathbf{p} &=& \left(n u_x, n u_y \right).
\end{eqnarray}

In the paraxial approximation, propagation in a homogeneous
medium of refractive index $n$ transforms the WDF according to the
relation
\begin{equation}
    \label{eq:paraxprop}
    W(\mathbf{q},\mathbf{p},z)=W(\mathbf{q}- \frac{z}{n} \mathbf{p},
    \mathbf{p},0).
\end{equation}
After the WDF has been propagated over a distance, the field
distribution can be recovered by \cite{Dragoman97, Bastiaans97}
\begin{equation}
 \label{eq:field}
 \varphi(\mathbf{q},z)\varphi^*(0,z)=\frac{1}{4 \pi^2}\int
 W(\mathbf{q}/2,\mathbf{p},z) \exp(i \mathbf{q}\mathbf{p})d
 \mathbf{p}.
\end{equation}
The field intensity distribution can also be found by
\begin{equation}
 \label{eq:intens}
 |\varphi(\mathbf{q},z)|^2= \frac{4 \pi^2}{k^2}\int W(\mathbf{q},\mathbf{p},z) d
 \mathbf{p}.
\end{equation}

Eqs. (\ref{eq:field}) and (\ref{eq:intens}) are all that is
needed for the evaluation of Fresnel diffraction fields. Consider
the diffraction pattern for a rectangular aperture in one
dimension illuminated by a monocromatic wave propagating in the
$z$ direction. The field distribution immediately after the
aperture is given by
\begin{eqnarray}
    \label{eq:fieldap}
    \varphi(x,0) = 1 ~~ &\rightarrow& ~~ |x|< l/2,
    \nonumber \\
    \varphi(x,0) = 0 ~~ &\rightarrow& ~~ |x|\geq l/2,
\end{eqnarray}
with $l$ being the aperture width.

Considering that $\varphi(x,0)$ is real we can write
\begin{equation}
    \label{eq:alter}
    \varphi\left(x+\frac{x'}{2}\right)
 \varphi^*\left(x-\frac{x'}{2}\right) =
 \mathrm{H} \left( \frac{l}{2}+\frac{x'}{2} - |x| \right)
 \mathrm{H} \left( \frac{l}{2}-\frac{x'}{2} - |x| \right).
\end{equation}

We then apply Eq.\ (\ref{eq:alter}) to the WDF definition Eq.\
(\ref{eq:Wigner}) to find
\begin{eqnarray}
   \label{eq:aperture}
   W(x,p_x)= 0 ~~& \rightarrow &~~ |x|\geq l/2, \nonumber \\
   W(x,p_x)=\frac{2\sin [k p_x (l - 2x)]}{k p_x} ~~&
    \rightarrow &~~ 0 \leq x < l/2, \nonumber \\
   W(x,p_x)=\frac{2\sin [k p_x (l + 2x)]}{k p_x} ~~&
    \rightarrow &~~ -l/2 \leq x \leq 0,
\end{eqnarray}
After propagation we obtain the following integral field
distribution
\begin{eqnarray}
 \label{eq:fresnel}
 |\varphi(x,z)|^2 &=&  \frac{4 \pi^2}{k^2}\left\{\int_{n( 2x-l)/(2z)}^{n x/z}
 \frac{2 \sin [k p_x (l - 2 z p_x/n - 2 x)]}{k p_x} d p_x \right . \nonumber\\
 && \left . + \int^{n(2x+l)/(2z)}_{n x/z} \frac{2 \sin [k p_x
 (l + 2 z p_x/n + 2 x)]}{k p_x} d p_x \right\}.
\end{eqnarray}

\begin{figure}[tbh]
    \centerline{\psfig{file=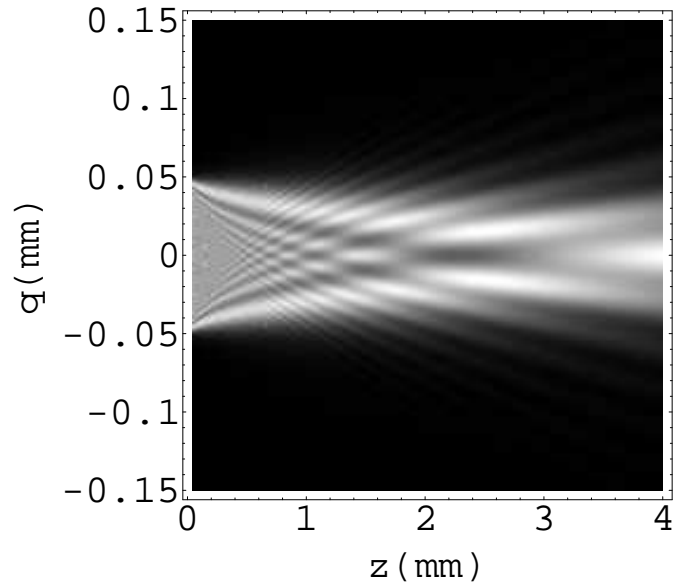, scale=1.5}}
    \caption{Fresnel diffraction pattern for a one-dimensional
aperture of width $0.1~\mathrm{mm}$ with $k=10^7$.}
\label{fig:fresnel}
\end{figure}

Fig.\ \ref{fig:fresnel} shows a typical diffraction pattern
obtained by numerical integration of Eq.\ (\ref{eq:fresnel}).

For wide angles paraxial approximation no longer applies and the
appropriate WDF transformation is now given by
\begin{eqnarray}
    \label{eq:wideshear}
    W(\mathbf{q},\mathbf{p},z)=W(\mathbf{q}- \frac{z \mathbf{p}}
    {\sqrt{n^2-|\mathbf{p}|^2}},
    \mathbf{p},0)~~&\rightarrow&~~ |\mathbf{p}|<n, \nonumber \\
    W(\mathbf{q},\mathbf{p},z)=0~~&\rightarrow&~~ \mathrm{otherwise}.
\end{eqnarray}

\begin{figure}[tbh]
   \parbox{80mm}{\centering{\psfig{file=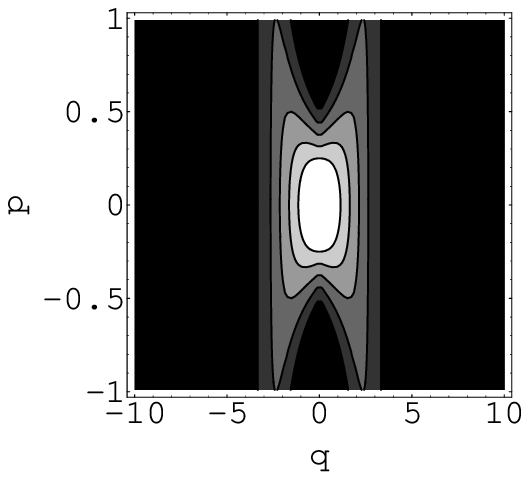,scale=1} \\ a)}}
    \parbox{80mm}{\centering{ \psfig{file=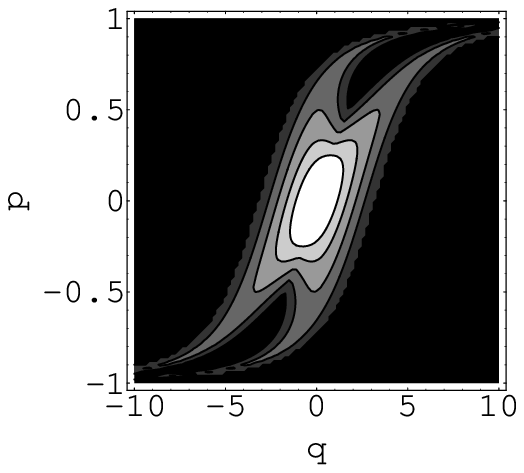, scale=1} \\ b)}}
    \caption{Propagation of the WDF in wide angle condition
($k=10^6~\mathrm{m}^{-1}$, horizontal scale in $\mu \mathrm{m}$).
a) Original distribution, b) after propagation over
$3~\mu\mathrm{m}$.}
 \label{fig:wideshear}
\end{figure}

Eq.\ (\ref{eq:wideshear}) shows that only the moments such that
$|\mathbf{p}|<n$ can be propagated \cite{Goodman68}. In fact, if
$|\mathbf{p}|/n=\sin \alpha$, with $\alpha$ the angle the ray
makes with the $z$ axis, It is obvious that the higher moments
would correspond to values of $|\sin \alpha|>1$; these moments
don't propagate and originate evanescent waves instead, Fig.\
\ref{fig:wideshear}. The net effect on near-field diffraction is
that the high-frequency detail near the aperture is quickly
reduced.

The field intensity can now be evaluated by the expression
\begin{eqnarray}
 \label{eq:wideangle}
 \frac{k^2|\varphi(x,z)|^2}{4 \pi^2} &=& \int_{p_1}^{p_0}{\frac{1}{k p_x}
  \sin \left\{k p_x \left[l - 2 \left(x-z p_x / \sqrt{n^2-p_x^2}\right)\right]\right\}
   d p_x }\nonumber\\
 && +  \int_{p_0}^{p_2}{\frac{1}{k p_x}
  \sin \left\{k p_x \left[l + 2 \left(x-z p_x / \sqrt{n^2-p_x^2}\right)\right]\right\}
   d p_x } ,
\end{eqnarray}
with
\begin{eqnarray}
    p_0 &=& \frac{n x}{\sqrt{x^2+z^2}}, \nonumber \\
    p_1 &=& \frac{n(2x-l)}{\sqrt{(2x-l)^2+4z^2}} , \nonumber \\
    p_2 &=& \frac{n(2x+l)}{\sqrt{(2x+l)^2+4z^2}} .
\end{eqnarray}

\begin{figure}[tbh]
    \centerline{\psfig{file=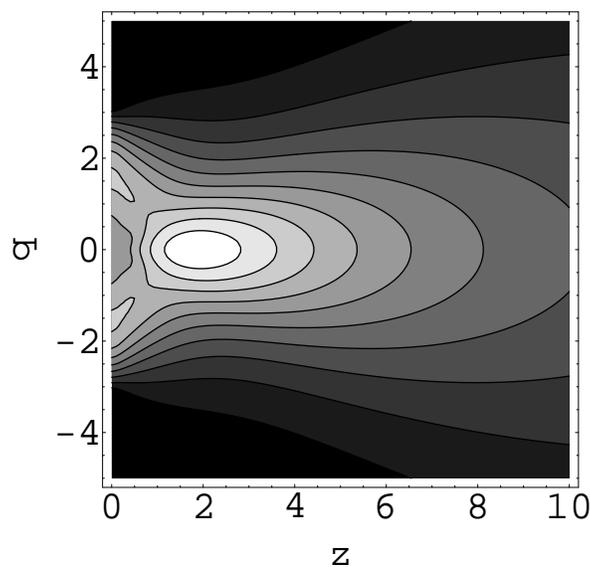, scale=1.5}}
    \caption{Near-field diffraction pattern when the aperture width
is exactly one wavelength; ($k=10^6~\mathrm{m}^{-1}$, both scales
in $\mathrm{\mu m}$).}
 \label{fig:wideangle}
\end{figure}

Fig.\ \ref{fig:wideangle} shows the near-field diffraction
pattern when the aperture is exactly one wavelength wide. The
situation is such that all the high frequencies appear at values
of $|p_x|>1$ and are evanescent, resulting in a field pattern with
one small minimum immediately after the aperture, after which the
beam takes a quasi-gaussian shape, without further minima. The
width of the sharp peak just after the aperture is considerably
smaller than one wavelength determining a super-resolution on
this region.
\section{Special relativity}
Special relativity deals with a 4-dimensional space-time endowed
with a pseudo-Euclidean metric which can be written as
\begin{equation}
    \label{eq:metric}
    d s^2 = dx^2 + dy^2 +dz^2 -dt^2,
\end{equation}
where the space-time is referred to the coordinates $(t, x, y,
z)$. Here the units were chosen so as to make $c=1$, $c$ being
the speed of light.

\begin{figure}[tbh]
    \centerline{\psfig{file=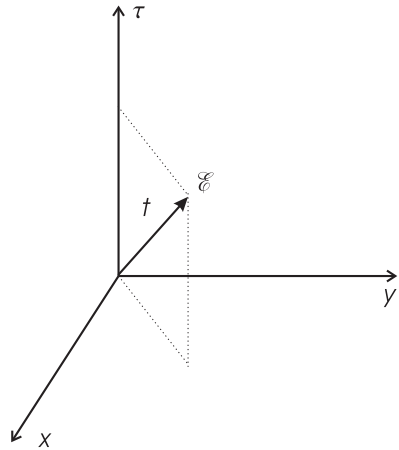, scale=0.4}}
    \caption{The relativistic frame in two dimensions. The $\tau$
coordinate is the \emph{proper time} while the distance to the
origin is the time measured in the frame at rest.}
\label{fig:coords}
\end{figure}

For a more adequate optical interpretation one can use
coordinates $(x,y,z, \tau)$ with $\tau$ the proper time
\cite{Inverno96}:
\begin{equation}
    \label{eq:tau}
    \tau = t \left(1- \frac{v^2}{c^2}\right)^{1/2} =
    \frac{t}{\gamma}~.
\end{equation}
Here $v^2=|\mathbf{v}|^2$, with $\mathbf{v}$
\begin{equation}
    \label{eq:speed}
    \mathbf{v} = \left(\frac{d x}{d t}, \frac{d y}{d t}, \frac{d z}{d t}\right),
\end{equation}
the usual 3-velocity vector. Fig.\ \ref{fig:coords} shows the
coordinates of an event $\mathcal{E}$ when just two spatial
coordinates are used together with coordinate $\tau$.

Considering Eqs.\ (\ref{eq:tau}, \ref{eq:speed}) and the new
coordinates the metric defined by Eq.\ (\ref{eq:metric}) becomes
\begin{equation}
 \label{eq:ds2}
 ds^2 = v^2 dt^2 -dt^2 = d\tau^2.
\end{equation}

\begin{figure}[tbh]
    \centerline{\psfig{file=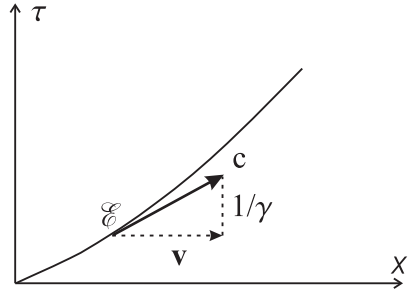, scale=0.5}}
    \caption{The curved line represents the word line of event
$\mathcal{E}$. The speed is represented by the vector
$\mathbf{v}$, which is the side of a rectangular triangle whose
hypotenuse has a magnitude $c$.}
 \label{fig:relspeed}
\end{figure}

The trajectory of an event in 4-space is known as its world line.
In Fig.\ \ref{fig:relspeed} we represent the world line of an
event $\mathcal{E}$ with coordinates $(x, \tau)$. At each
position the derivative of the world line with respect to $\tau$
is
\begin{equation}
    \label{eq:xprime}
    \dot{x} = \frac{d x}{d t}~ \frac{d t}{d \tau} = \gamma v_x.
\end{equation}
More generally we can write
\begin{equation}
    \label{eq:speedmu}
    \overrightarrow{\nu} = \gamma \mathbf{v},
\end{equation}
with
\begin{equation}
    \label{eq:mu}
    \overrightarrow{\nu} = \left(\dot{x}, \dot{y}, \dot{z}\right).
\end{equation}

It follows from Eq.\ (\ref{eq:mu}) that, at each point, the
components of $\mathbf{v}$ are the direction cosines of the
tangent vector to the world line through that point.

We must now state a basic principle equivalent to Fermat's
principle for ray propagation in optics given in Eq.\
(\ref{eq:Fermat}). Taking into consideration the relativistic
Lagrangian we can state the following variational principle
\cite{Goldstein80, Jose98}:
\begin{equation}
    \label{eq:relprinciple}
    \delta \int {\left(m \gamma - V \right) d s} = 0,
\end{equation}
where $m$ is the rest mass of the particle and $V$ is the local
potential energy. It can be shown that $m \gamma$ represents the
kinetic energy of the particle. Using Eq.\ (\ref{eq:ds2}) a
straightforward calculation shows that $ds=d \tau$ so that we can
also write
\begin{equation}
    \label{eq:relprinciple2}
    \delta \int {\left(m \gamma - V \right) d \tau} = 0.
\end{equation}
The new principle means that the particle will choose a trajectory
moving away from high potential energies much as the optical ray
moves away from low refractive index. From this principle we can
derive all the equations of relativistic mechanics, in a similar
manner as from Fermat's principle it is possible to derive light
propagation.

Comparing this equation with Eq.\ (\ref{eq:Lagrangian}), we can
define the Lagrangian of the mechanical system as
\begin{equation}
    \label{eq:rellag}
    L = \left(m \gamma -V \right).
\end{equation}

We now have a 4-dimensional space, while in the Hamiltonian
formulation of optics we used 3 dimensions. The following list
shows the relationship between the two systems, where we refer
first to the relativistic system and then to the optical system.
\begin{eqnarray*}
   x & \leftrightarrow & x, \\
   y & \leftrightarrow & y, \\
   z & \leftrightarrow & \mathrm{no ~ equivalent}, \\
   \tau & \leftrightarrow & z. \
\end{eqnarray*}

In the mapping from optical propagation to special relativity the
optical axis becomes the \emph{proper time} axis and the ray
direction cosines correspond to the components of the speed vector
$\mathbf{v}$. The refractive index has no direct homologous; it
will be shown that in special relativity we must consider a
non-homogeneous medium with different refractive indices in the
spatial and \emph{proper time} directions. We can derive the
 conjugate momentum from the Lagrangian using the standard procedure:
\begin{equation}
    \label{eq:momentum}
    \mathbf{p} = \left( \frac{m \overrightarrow{\nu}}{\gamma} \right) = m \mathbf{v}.
\end{equation}
Comparing with Eq.\ (\ref{eq:momenta2}) it is clear that $m$ is
the analogous of the position dependent refractive index.

The system Hamiltonian can be calculated
\begin{eqnarray}
    \label{eq:relham}
    H &=& \mathbf{p} \cdot \mathbf{v} -L \nonumber \\
      &=& \frac{m \nu^2}{\gamma } -L \nonumber \\
      &=& -\frac{m}{\gamma}+ V  .
\end{eqnarray}

The canonical equations follow directly from Eq.\
(\ref{eq:canonical})
\begin{eqnarray}
    \label{eq:relcanonical}
    \overrightarrow{\nu}&=& \gamma \mathbf{v}, \nonumber \\
    \frac{d \mathbf{p}}{d \tau}
    &=& - \mathrm{grad}H,
\end{eqnarray}
where the gradient is taken over the spatial coordinates, as
usual.

The first of the equations above is the same as Eq.\
(\ref{eq:mu}), while the second one can be developed as
\begin{equation}
    \label{eq:canon2}
    \frac{d}{d \tau} \left( \frac{m}{\gamma }
    \overrightarrow{\nu}\right)
     = \mathrm{grad}\left(\frac{m}{\gamma} - V \right).
\end{equation}
Considering that from the quantities inside the gradient only the
potential energy should be a function of the spatial coordinates,
we can simplify the second member:
\begin{eqnarray}
    \label{eq:canon3}
    \frac{d}{d \tau} \left( \frac{m}{\gamma }
    \overrightarrow{\nu}\right)
     &=& -\mathrm{grad}V \\
     \label{eq:canon3b}
    \frac{d}{d \tau} \left(m \mathbf{v} \right)
    &=&  -\mathrm{grad}V.
\end{eqnarray}
Eq.\ (\ref{eq:canon3b}) is formally equivalent to the last two
Eqs.\ (\ref{eq:canonicaldiff}), confirming that the conjugate
momentum components are proportional to world line's direction
cosines in 4-space. The total refractive index analogue can now
be found to be $m -V / \gamma$. We can check the validity of Eq.\
(\ref{eq:canon3b}) by replacing the $\tau$ derivative by a
derivative with respect to $t$.
\begin{equation}
    \label{eq:canon4}
    \frac{d}{d t} \left( m \gamma
    \mathbf{v}\right)
     = -\mathrm{grad}V,
\end{equation}
where $m \gamma$ is the relativistic mass and the product $m
\gamma \mathbf{v}$ is the relativistic momentum.

If the mass is allowed to be coordinate dependent, as a mass
distribution through the Universe, the passage between Eqs.\
(\ref{eq:canon2}) and (\ref{eq:canon3}) is illegitimate and we
are led to equations similar to Eq.\ (\ref{eq:canonicaldiff}).
The consideration of a coordinate dependent mass allows the
prediction of worm tubes, the analogues of optical waveguides,
and black holes, which would have optical analogues in high
refractive index glass beads with gradual transition to vacuum
refractive index.
\section{De Broglie's wavelength}
The formal equivalence between light propagation and special
relativity in the $x, y, z, \tau$ frame suggests that de Broglie's
wavelength may be the formal equivalent of light wavelength. We
would like to associate an event's world line to a light ray and
similarly we want to say that, in the absence of a potential, the
event's world line is the normal to the wavefront at every point
in 4-space. We must start from a basic principle, stating that
each particle has an intrinsic frequency related to it's mass in a
similar way as the wavelength of light is related to the
refractive index; we state this principle by the equation
\begin{equation}
    \label{eq:intfreq}
    f = \frac{m}{h},
\end{equation}
where $h$ is Planck's constant. If we remember that everything is
normalized to the speed of light by $c=1$, Eq.\
(\ref{eq:intfreq}) is equivalent to a photon's energy equation
\begin{equation}
    E = h f.
\end{equation}
So we have extended an existing principle to state that any
particle has an intrinsic frequency that is the result of
dividing it's equivalent energy $E = m c^2$ by Planck's constant.

In the 4-space $x, y, z, \tau$ frame a material particle travels
in a trajectory with direction cosines given by the components of
$\mathbf{v}$ and consistently a photon travels along a spatial
direction with zero component in the $\tau$ direction. The
intrinsic frequency defined by Eq.\ (\ref{eq:intfreq}) originates
a wavelength along the world line given by
\begin{equation}
    \label{eq:worldlength}
    \lambda_w = \frac{c}{f}= \frac{h}{m c},
\end{equation}
where we have temporarily removed the $c$ normalization for
clarity reasons.

\begin{figure}[tbh]
    \centerline{\psfig{file=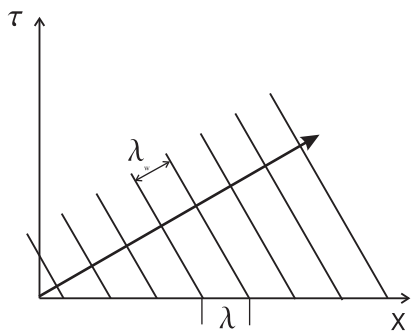, scale=0.5}}
    \caption{The moving particle has a world wavelength
$\lambda_w=h/(mc)$ and a spatial wavelength $\lambda=h/(mv)$.}
\label{fig:broglie}
\end{figure}

As shown in Fig.\ \ref{fig:broglie}, when projected onto 3-space
$\lambda_w$ defines a spatial wavelength such that
\begin{equation}
    \label{eq:broglie}
    \lambda = \frac{\lambda_w c}{v} = \frac{h}{ p}.
\end{equation}
The previous equation defines a spatial wavelength which is the
same as was originally proposed by de Broglie in the
non-relativistic limit. When the speed of light is approached
Eq.\ (\ref{eq:broglie}) will produce a wavelength $\lambda
\rightarrow \lambda_w$ while de Broglie's predictions use the
relativistic momentum and so $\lambda \rightarrow \infty$ when $v
\rightarrow c$.
\section{Wave propagation and Schr\"odinger equation}
The arguments in the previous paragraphs lead us to state the
general principle that a particle has an associated frequency
given by Eq.\ (\ref{eq:intfreq}) and travels on a world line
through 4-space with the speed of light. In a generalization to
wave packets, we will respect the formal similarity with light
propagation and state that all waves travel in 4-space with the
speed of light. A particle about which we know mass and speed but
know nothing about its position will be represented by a
monocromatic wave and a moving particle in general will be
represented by a wave packet.

According to general optical practice we will say that the field
must verify the wave equation
\begin{equation}
    \label{eq:waveeq}
    \left(\nabla^2 - \frac{\partial^2 }{\partial
    t^2}\right) \varphi(\mathcal{P}, t) =0,
\end{equation}
where $\mathcal{P}$ is a point in 4-space and $\nabla^2$ is an
extended laplacian operator
\begin{equation}
    \nabla^2 = \frac{\partial^2}{\partial x^2} +\frac{\partial^2}{\partial y^2}
     +\frac{\partial^2}{\partial z^2} +\frac{\partial^2}{\partial
     \tau^2}.
\end{equation}
In Eq.\ (\ref{eq:waveeq}) we have returned to the $c=1$
normalization in order to treat all the coordinates on an equal
footing. Due to the special 4-space metric we will assume that
$\varphi(\mathcal{P},t)$ is of the form
\begin{equation}
    \label{eq:phipt}
    \varphi(\mathcal{P},t) = \Phi(\mathcal{P}) e^{i 2 \pi f t},
\end{equation}
with $f$ given by Eq.\ (\ref{eq:intfreq}). Notice that we used a
plus sign in the exponent instead of the minus sign used in
optical propagation; this is due to the special 4-space metric.

Not surprisingly we will find that, in the absence of a potential,
Eq.\ (\ref{eq:waveeq}) can be written in the form of Helmoltz
equation
\begin{equation}
    \label{eq:Helmoltz}
    \left(\nabla^2 + k^2 \right)\Phi(\mathcal{P})=0,
\end{equation}
with
\begin{equation}
    \label{eq:kapa}
    k = \frac{2 \pi}{\lambda_w}.
\end{equation}

If we take into consideration Eq.\ (\ref{eq:tau}), the laplacian
becomes
\begin{eqnarray}
    \label{eq:laplacian}
    \nabla^2 &=& \frac{\partial^2}{\partial x^2} +\frac{\partial^2}{\partial y^2}
     +\frac{\partial^2}{\partial z^2} +\gamma^2 \frac{\partial^2}{\partial
     t^2}\nonumber \\
     &=& \nabla_3^2 +\gamma^2 \frac{\partial^2}{\partial t^2},
\end{eqnarray}
where $\nabla_3^2$ represents the usual laplacian operator in
3-space.

In order to derive Schr\"odinger equation we re-write Eq.\
(\ref{eq:waveeq})
\begin{equation}
    \label{eq:waveeq2}
    \nabla^2 \varphi(\mathcal{P}, t) + i k
    \frac{\partial \varphi(\mathcal{P}, t) }{\partial
    t} =0,
\end{equation}
and using Eq.\ (\ref{eq:laplacian})
\begin{equation}
    \label{eq:waveeq3}
    \nabla_3^2 \varphi(\mathcal{P}, t) + i k(\gamma^2+1)~
    \frac{\partial \varphi(\mathcal{P}, t) }{\partial
    t} = 0.
\end{equation}

In a non-relativistic situation $\gamma\rightarrow 1$.
Considering Eq.\ (\ref{eq:worldlength}) we can write Eq.\
(\ref{eq:waveeq3}) in the form of Schr\"odinger equation
\cite{Gasiorowicz96}
\begin{equation}
    \label{eq:Schrod}
    i \hbar \frac{\partial \varphi(\mathcal{P}, t)}{\partial t} =
    - \frac {\hbar^2}{2 m}\nabla_3^2 \varphi(\mathcal{P}, t),
\end{equation}
where $\hbar = h / (2 \pi)$.

Eq.\ (\ref{eq:Schrod}) retains the symbol $\mathcal{P}$
representing a point in 4-space. It must be noted, though, that
in a non-relativistic situation $\tau \rightarrow t$ and we can
say that $\varphi(\mathcal{P}, t) \rightarrow \varphi(P, t)$ with
$P$ having the 3-space coordinates of $\mathcal{P}$.

In the presence of a potential we have to consider that momentum
is no longer preserved, as shown by Eq.\ (\ref{eq:canon3b}). This
can be taken into account when we evaluate the laplacian in Eq.\
(\ref{eq:waveeq3}) by the inclusion of an extra term $-V
\varphi(\mathcal{P},t)$. We will the end up with the
Schr\"odinger equation in a potential.
\section{Heisenberg's uncertainty principle}
For the pair of associated variables $x$ and $p_x$, Heisenberg's
uncertainty principle states that there is an uncertainty governed
by the relation
\begin{equation}
    \label{eq:heisenberg}
    \Delta x \Delta p_x \geq \frac{\hbar}{2}.
\end{equation}

The interpretation of the uncertainty principle is that the best
we know a particle's position, the least we know about its
momentum and vice-versa; the product of the position and momentum
distribution widths is a small number multiplied by Planck's
constant. An application of the uncertainty relationship is
usually found in the diffraction of a particle by an aperture.

If we assume that the localization of a particle with momentum
$p_x$ can be done with an aperture of width $\Delta x$, we can use
Fraunhofer diffraction theory to say that further along the way
the probability of finding the particle with any particular value
of it's momentum will be given by the square of the aperture
Fourier transform, considering de Broglie's relationship for the
translation of momentum into wavelength.

A rectangular aperture of width $\Delta x$ has a Fourier
transform given by
\begin{equation}
    \label{eq:fourierap}
    A (f_x) = \Delta x \mathrm{~sinc}(\Delta x f_x),
\end{equation}
where $\mathrm{sinc}$ is the usual $\sin(x)/x$ function.

Considering de Broglie's relationship given by Eq.\
(\ref{eq:broglie}), making $f_x = 1/\lambda$ and the fact that the
Fourier transform must be squared we can write
\begin{equation}
    \label{eq:apertmoment}
    P (p_x) = \Delta x^2 \mathrm{~sinc}^2 \left(\frac{\Delta x
    p_x}{h} \right).
\end{equation}
The second member on the previous equation has its first minimum
for $\Delta x p_x /h = \pi$ and so we can say that the spread in
momentum is governed by $\Delta x \Delta p_x = \pi h$ and Eq.\
(\ref{eq:heisenberg}) is verified.

If we accept that wave packets propagate in 4-space at the speed
of light and that the momentum is given by Eq.\
(\ref{eq:momentum}), there is a upper limit to the modulus of the
momentum $|\mathbf{p}|\leq m c$. In the propagation of light rays
we found a similar limitation as $|\mathbf{p}|\leq n$ with the
results in light diffraction exemplified by Eq.\
(\ref{eq:wideangle}) and Fig.\ \ref{fig:wideangle}. It is
expected that the same effects will be present in particle
diffraction and in fact Figs.\ \ref{fig:wideshear} and
\ref{fig:wideangle} could also represent the diffraction of a
stationary particle by an aperture with width equal to
$\lambda_w$. The strong peak about half one wavelength in front
of the aperture shows that the particle is localized in a region
considerably smaller than its wavelength and, above all, shows
the inexistence of higher order peaks.
\section{Conclusions}
Special relativity was shown to be formally equivalent to light
propagation, provided the time axis is replaced by the
\emph{proper time}. In this coordinate set all particles follow w
world line at the speed of light and can be assumed to have an
intrinsic frequency given by $m c^2/h$. Quantum mechanics is then
a projection of 4-space wave propagation into 3-space. Important
conclusions were possible through the analogy with light
propagation and diffraction.

It was possible to derive Schr\"odinger equation and it was shown
that Heisenberg's uncertainty principle may be violated in
special cases in the very close range, similarly to what had
already been shown to happen in light diffraction
\cite{Almeida00:3}.

Future work will probably allow the derivation of relativistic
quantum mechanics conclusions, through the use of the Wigner
Distribution Function for the prediction of wave packet
propagation in 4-space.
\section{Acknowledgements}
The author acknowledges the many fruitful discussions with
Estelita Vaz from the Mathematics Department of Universidade do
Minho, especially on the subject of relativity.
%
  \bibliography{aberrations}   
  \bibliographystyle{OSA}   

\end{document}